\documentclass[12pt,preprint]{aastex}

\usepackage{times}
\usepackage{mathptm}
\usepackage{graphicx}
\usepackage{dcolumn}
\usepackage{bm}
\usepackage{psfig}

\begin{document}


\title{Constrained Transport Algorithms for Numerical Relativity.  \\
I.  Development of a Finite Difference Scheme}


\author{David L. Meier}
\affil{Jet Propulsion Laboratory, California Institute of Technology, 
    Pasadena, CA 91109}



\begin{abstract}
A scheme is presented for accurately propagating the gravitational field
constraints in finite difference implementations of numerical relativity.
The method is based on similar techniques used in astrophysical
magnetohydrodynamics and engineering electromagnetics, and has properties
of a finite differential calculus on a four-dimensional manifold.  It is
motivated by the arguments that 1) an evolutionary scheme that naturally
satisfies the Bianchi identities will propagate the constraints, and 2)
methods in which temporal and spatial derivatives commute will satisfy
the Bianchi identities implicitly.  The proposed algorithm exactly
propagates the constraints in a local Riemann normal coordinate system;
{\it i.e.}, all terms in the Bianchi identities (which all vary as
$\partial^3 g$) cancel to machine roundoff accuracy at each time step.
In a general coordinate basis, these terms, and those that vary as
$\partial g~\partial^2 g$, also can be made to cancel, but differences of
connection terms, proportional to $(\partial g)^3$, will remain, resulting
in a net truncation error.  Detailed and complex numerical experiments
with four-dimensional staggered grids will be needed to completely
examine the stability and convergence properties of this method.

If such techniques are successful for finite difference implementations
of numerical relativity, other implementations, such as finite element
(and eventually pseudo-spectral) techniques, might benefit from schemes that use
four-dimensional grids and that have temporal and spatial derivatives
that commute.
\end{abstract}





\section{Introduction}

The quest for solutions of dynamical strong gravity problems, such as
black hole formation or mergers and gamma-ray burst production, that are
astrophysically relevant and accurate enough to predict gravitational
wave forms, has occupied much of the last half of the $20^{\rm th}$
century and beyond.  Its history has been frustrated by the need to
address several unforeseen numerical problems, including 1) proper
initial data to begin the evolution, 2) the development of coordinate
and physical singularities during the evolution, and 3) the growth of
numerical instabilities in the time-dependent solutions.  These problems
have been dealt with by addressing each in turn with such techniques as
1) elevation of the initial data problem to a complete sub-field; 2)
development and analysis of appropriate gauge/slicing conditions that
avoid coordinate singularities; 3) excision of black hole centers,
inside horizons, to avoid physical singularities; and 4) the use of
symmetric/hyperbolic equations to enhance numerical stability.

One of the several remaining problems in this field is that, for some
problems and some coordinate systems, the constraint-violating modes
will grow exponentially.  These eventually overwhelm any solution in
only a short period of time ($< 100 \, M$), rendering a long simulation
(and the computation of any gravitational wave forms) impossible.
For very high-resolution simulations the exponential growth of errors
begins early and continues until the errors diverge.  On the other
hand, for simulations with coarse spatial resolution the errors begin
and remain at the truncation level until the much smaller constraint
violations grow to a level that exceeds the truncation accuracy.
Then the solution joins the general exponential growth seen in the
high-resolution simulations, blowing up in much the same manner as
in the high-resolution case \cite{scheel02}.  The similar behavior of
this error at a variety of mesh spacings indicates that it may be a
numerical solution to the discrete equations that are being integrated.
Current attempts to solve this problem include adding the constraints
as penalty functions to the evolution equations and techniques that
re-converge the constraint equations every few time steps.

Constraint propagation is also an issue in the solution of Maxwell's
equations.  Techniques for doing so in the fields of astrophysical
magnetohydrodynamics (MHD) and electromagnetics of antennas and waves
have been in place for decades.  These enforce the constraints not just
stably, but {\em to machine accuracy}.  Finite difference methods for
constraint propagation in astrophysical MHD are known as the Evans-Hawley
constrained transport (CT) method \cite{eh88} and now are an integral
part of publicly used codes, such as ZEUS \cite{sn92a,sn92b} and ZEUS-3D 
\cite{clarke96}. In engineering
electromagnetics these are known as the Yee algorithm \cite{yee66},
and have many variants \cite{deraedt02}.  They all involve building a mesh
that is staggered in space, and often in time as well, and then defining
appropriate vector and scalar quantities at whole or half mesh points.

While the success of CT for electromagnetics is certainly due in part
to the linearity of the physical equations, its ability to maintain
accuracy of the solenoidal (${\bm \nabla} \cdot {\bm B} = 0$) and Coulomb
(${\bm \nabla} \cdot {\bm E} = 4 \pi \rho_c$) constraints, to a few
parts in $10^{14-15}$ over tens or hundreds of thousands of time steps,
is enticing.  If such an algorithm can be found for numerical relativity,
it could be as useful as excision and other such proven methods.

In this paper the properties of the electromagnetic CT method (and of
spacetime itself) are examined, and a similar method is developed for
numerical relativity.  It is the thesis of this paper, and subsequent
ones in this series, that CT methods work because they build spacetimes
in which the temporal and spatial derivatives commute ($[\partial_0,
\partial_i] = O(\epsilon_r)$, where $\epsilon_r$ is the machine roundoff
error).  This naturally enforces the Bianchi identities, and it is those
implicit identities that propagate the constraints.  This thesis is not
fully tested in this first paper on the subject.  In fact, it will take
some time and numerical effort to verify or refute it.  Instead, only the
first steps are taken here.  It is shown that the proposed CT scheme for
NR works exactly in Riemann normal coordinates; in general coordinates
most terms also cancel.  Detailed numerical simulations in four dimensions
will be needed to fully explore the method's stability properties to see
if these conditions are sufficient for stable constraint propagation.
If successful, however, similar CT methods should be possible for other 
implementations of numerical relativity, not just for finite differences.

This paper is intended to be the first in a long series that will
culminate in a numerical code that is capable of simulating black hole
formation and the gamma-ray burst jet generation and gravitational wave
production that is expected to result from such events.  These issues are
important both for dealing properly with the energetics in the system
as well as with the expected observational consequences of the event.
In order to treat these problems properly, such a code must be capable
of evolving the relativistic gravitational field, as well as the
fluid matter flowing within that field {\em and} the electromagnetic
field generated by currents flowing in that matter.  Part of achieving
this goal will be to present a consistent numerical method, from the
relativistic gravitational field to specific issues of stellar mergers
and collapse.  In this paper we lay the groundwork for generating the
time-dependent gravitational field and for evolving the electromagnetic
field in that metric.  In subsequent papers we will present tests of the
constrained transport techniques developed herein, and eventually add
the stress-energy due to matter and fluid motion to complete the code.
Then, specific astrophysical problems will be addressed.  The ultimate
aim of this work is to foster a closer relationship between astronomers
who {\em observe} black hole systems and those numerical physicists and
astrophysicists who study them {\em theoretically}.

\section{Review of CT for Electromagnetics in Flat Spacetime}

\begin{figure}
\plotone{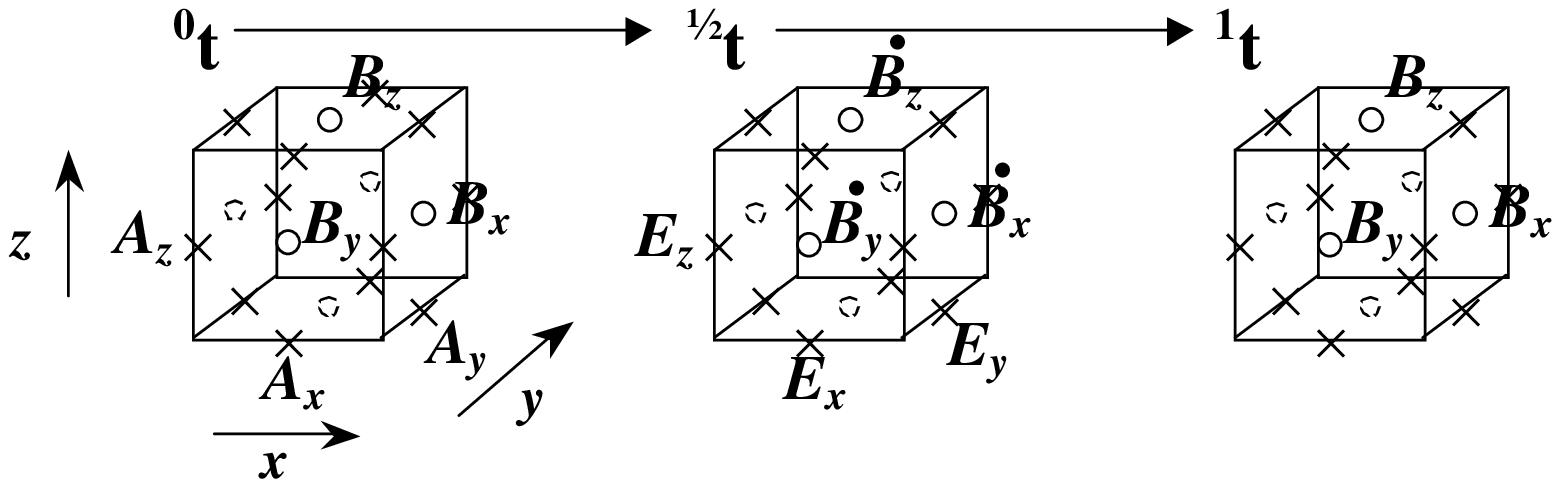}
\caption{Space-time representation for the Evans-Hawley CT scheme.
Open circles are face-centered on the cubes and crosses are edge-centered.
\label{fig1}}
\end{figure}

\subsection{The Evans-Hawley CT Method for MHD}

For astrophysical MHD the field equations that are solved are
\begin{eqnarray}
\label{mhd_evolution}
{\dot {\bm B}} & = & -c ~ {\bm \nabla} \times {\bm E}
\end{eqnarray}
with the solenoidal constraint
\begin{eqnarray}
\label{mhd_constraint}
\nabla \cdot {\bm B} & = & 0
\end{eqnarray}
(Additional equations are solved, of course, including the conservation
of mass, momentum, and energy; but these are not relevant here.)
The Evans-Hawley constrained transport technique satisfies equation
(\ref{mhd_constraint}) on the initial hypersurface, usually to
machine accuracy, and then uses a differencing scheme for equation
(\ref{mhd_evolution}) that ensures that equation (\ref{mhd_constraint})
is satisfied on each subsequent hypersurface to the same level of accuracy
as on the first.  This is done by staggering the grid in space and time
(see Figure \ref{fig1}).  At whole time steps the magnetic field vector
components are defined normal to, and centered on, grid cube faces.
At half time steps the electric vector is defined parallel to, and
centered on, cube edges.  For the initial conditions the magnetic field
is derived from a vector potential
\begin{eqnarray}
\label{initial_b}
{\bm B} & = & \nabla \times {\bm A}
\end{eqnarray}
This vector potential is defined on cube edges at $~t = ~ ^0t$, and
for evolutionary computations ${\dot {\bm B}} \equiv \partial {\bm B} /
\partial t$ is defined on cube faces, centered in time between $~ ^n{\bm
B}$ and $~ ^{n+1}{\bm B}$, {\it i.e.}, at $~t =~ ^{n+\frac{1}{2}}t$.

At $~t =~ ^0t$ we see that the solenoidal constraint is satisfied to
machine accuracy by this method.  For $\Delta x = \Delta y = \Delta z$
we have, simply,
\begin{eqnarray}
\nabla \cdot {\bm B} & = & ~ _+B_x ~ - ~ _-B_x ~ + ~ _+B_y ~ - ~ _-B_y ~ + ~ _+B_z ~ - ~ _-B_z
\nonumber 
\\
& = & \nabla \cdot \nabla \times {\bm A} 
\nonumber 
\\
& = & (_{++}A_z \, - \, _{+-}A_z) ~ - ~ (_{++}A_y \, - \, _{+-}A_y) ~ - ~ 
      (_{-+}A_z \, - \, _{--}A_z) ~ + ~ (_{-+}A_y \, - \, _{--}A_y) ~ + ~ 
\nonumber 
\\
& &   (_{++}A_x \, - \, _{+-}A_x) ~ - ~ (_{++}A_z \, - \, _{-+}A_z) ~ - ~ 
      (_{-+}A_x \, - \, _{--}A_x) ~ + ~ (_{+-}A_z \, - \, _{--}A_z) ~ + ~ 
\nonumber 
\\
& &   (_{++}A_y \, - \, _{-+}A_y) ~ - ~ (_{++}A_x \, - \, _{-+}A_x) ~ - ~ 
      (_{+-}A_y \, - \, _{--}A_y) ~ + ~ (_{+-}A_x \, - \, _{--}A_x) 
\nonumber 
\\
\label{divb_discrete}
& = & O(\epsilon_r) 
\end{eqnarray}
where $~_+ B_i$ and $~_- B_i$ are components on upper and lower cube
$i$-faces, respectively, and the two pre-appended subscript signs on
vector potential components $A_j$ indicate the $j^{\rm th}$ edge at the
intersection of the upper and/or lower cube faces with normals in the two
spatial dimensions orthogonal to $j$.  Similarly, taking the numerical
divergence of equation (\ref{mhd_evolution}) at $~ ^{\frac{1}{2}}t$,
we have
\begin{eqnarray}
\nabla \cdot {\dot {\bm B}} & = & -c ~ \nabla \cdot \nabla \times {\bm E} ~ = ~ O( \epsilon_r )
\nonumber
\end{eqnarray}
so, at time $~t =~ ^1t$ the magnetic field remains divergence-free
\begin{eqnarray}
\nabla \cdot ~ ^1{\bm B} & = & \nabla \cdot ~ ^0{\bm B} ~ + 
~ \Delta t ~ ~ ^{\frac{1}{2}}(\nabla \cdot {\dot {\bm B}}) ~ = O(\epsilon_r)
\nonumber
\end{eqnarray}
because it is the sum of two divergence-free fields.  The solenoidal
constraint is therefore preserved to machine accuracy {\em specifically
because the vector identity} $\nabla \cdot \nabla \times {\bm E} = 0$
{\em is naturally satisfied by the differencing scheme}.

In the above equations the electric field can be computed by any means,
and CT still would be maintained.  Ideal MHD codes use Ohm's law with
infinite conductivity to relate the fluid velocity and the magnetic field:
\begin{eqnarray}
{\bm E} & = & -\frac{{\bm v}}{c} \times {\bm B}
\nonumber
\end{eqnarray}
Some astrophysical codes, like ZEUS and several in Japan, use the method
of characteristics to preserve Alfven waves in the ${\bm E}$ update, and
so are often called MOC-CT codes.  However, it is important to realize
that CT results simply from the grid staggering and has nothing to do
with the method of characteristics or any other $\bm E$ update scheme.

\subsection{The Yee Algorithm for Electromagnetics}

\begin{figure}
\plotone{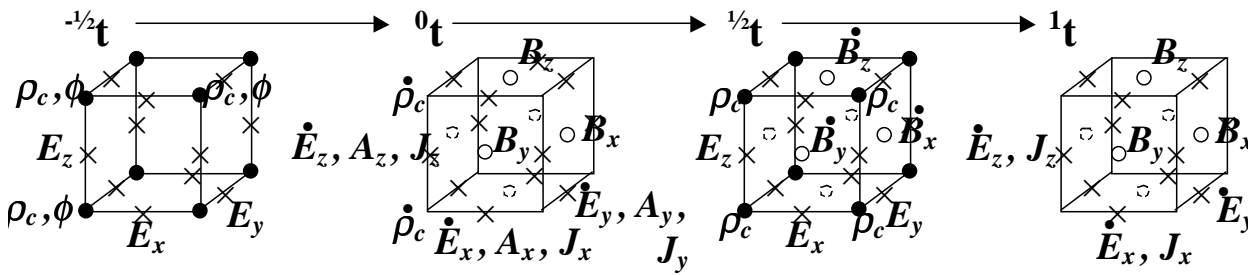}
\caption{Space-time representation for the Yee algorithm.  Similar
to Figure \ref{fig1}, but with filled circles located at cube
corners. Another time step has been added ($^{-\frac{1}{2}}t$, along
with electric field quantities.
\label{fig2}}
\end{figure}

In full electrodynamics two more Maxwell's equations are used to determine
the electric field, instead of Ohm's law.  In vacuum these are
\begin{eqnarray}
\label{ed_evolution}
{\dot {\bm E}} & = & c ~ \nabla \times {\bm B} ~ - ~ 4 \pi {\bm J} 
\\
\nabla \cdot {\bm E} & = & 4 \pi \rho_c
\label{ed_constraint}
\end{eqnarray}
Figure \ref{fig1} then must be modified to include the new quantities
${\bm J}$ (current density), $\rho_c$ (charge density), and $\phi$
(electric potential).  (See Figure \ref{fig2}.)  We also add another
initial data problem at $~t =~ ^{-\frac{1}{2}}t$ that solves equation
(\ref{ed_constraint}) for ${\bm E}$.  If, for the moment, we choose 
the Coulomb gauge and ignore the vector potential, we have
\begin{eqnarray}
\nabla^2 \phi & = & - 4 \pi \rho_c 
\nonumber
\\
{\bm E} & = & - \nabla \phi
\nonumber
\end{eqnarray}
Electric potential and charge density then must be defined at cube corners
on half time steps, and current density must be defined on cube edges
at whole time steps (same as $curl \, {\bm B}$, ${\bm A}$, and ${\dot
{\bm E}}$; see Figure \ref{fig2}).  In an arbitrary gauge, we will have
\begin{eqnarray}
\label{ed_gauge}
{\bm E} & = & -\nabla \phi ~ - ~ \frac{1}{c} {\dot {\bm A}}
\end{eqnarray}
indicating that $\dot{\bm A}$ and ${\bm E}$ are co-located.  The equations
are closed by specifying the evolution of $\rho_c$ and ${\bm J}$, which
together must satisfy the conservation of charge
\begin{eqnarray}
\nabla \cdot {\bm J} & = & - \dot{\rho}_c
\end{eqnarray}
so $\dot{\rho}_c$ must be defined on cell corners at whole time steps.
This method of staggering was suggested buy Yee almost 40 years ago
\cite{yee66}.

Constraints are preserved in the Yee algorithm in the same manner as
in the Evans-Hawley algorithm.  For Faraday's law and the solenoidal
constraint, the procedure is identical.  And for Ampere's law we have
\begin{eqnarray}
\nabla \cdot {\dot {\bm E}} & = & c ~ \nabla \cdot \nabla \times {\bm B} ~ - ~ 4 \pi \nabla \cdot {\bm J}
\nonumber 
\\
& = & O(\epsilon_r) ~ + ~ 4 \pi \dot{\rho}_c
\nonumber 
\end{eqnarray}
which gives the following update for $div \, {\bm E}$:
\begin{eqnarray}
~ ^{\frac{1}{2}}(\nabla \cdot {\bm E}) & = & ~ ^{-\frac{1}{2}}(\nabla \cdot {\bm E}) ~ + ~ ^1(\nabla \cdot \dot{{\bm E}}) ~ \Delta t
\nonumber 
\\
& = & 4 \pi ~ ^{-\frac{1}{2}}\rho_c ~ + ~ 4 \pi ~ \Delta t ~ ~ ^1\dot{\rho}_c ~ + ~ O(\epsilon_r) 
\nonumber 
\\
& = & 4 \pi ~ ^{\frac{1}{2}}\rho_c~ + ~ O(\epsilon_r) 
\nonumber 
\end{eqnarray}
So the Coulomb constraint is preserved to machine accuracy as long as
$\rho_c$ is conserved in the update.

\subsection{Covariant formulation of CT for Electrodynamics}

\begin{figure}
\centerline{\psfig{figure=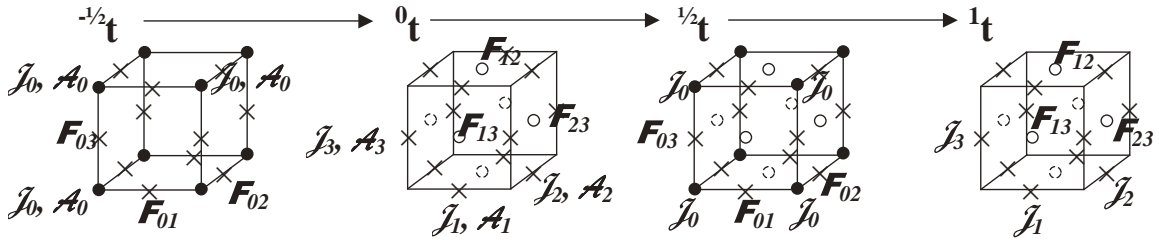,height=1.0in}}
\caption{
Same as Figure \ref{fig2}, but with covariant notation.  Note placement
of vector components (on hypercube edges) and Faraday tensor components
(on hypercube faces).
\label{fig3}}
\end{figure}

Figure \ref{fig3} re-casts the Yee algorithm in covariant form, using the
Faraday tensor ${\bf F}$ (instead of the vector fields), the four-current
${\bf J}$, and the vector four-potential ${\bf A}$.  This will give
important clues to developing a CT scheme for Einstein's field equations.
Maxwell's equations then become, including constraints,
\begin{eqnarray}
\label{ed_cov_magnetic}
\nabla \cdot {\bf M} & = & 0 
\\
\nabla \cdot {\bf F} & = & 4 \pi {\bf J}
\label{ed_cov_electric}
\end{eqnarray}
where ${\bf M} \equiv ~ ^*{\bf F}$ is the Maxwell tensor (the dual
of ${\bf F}$) and $\nabla$ is now the {\em four}-gradient operator
$\nabla \equiv (\frac{\partial}{\partial t}, \frac{\partial}{\partial
x}, \frac{\partial}{\partial y}, \frac{\partial}{\partial z})$.  Because
${\bf M}$ and ${\bf F}$ are antisymmetric, they satisfy tensor identities
(analogous to the vector identities $\nabla \cdot \nabla \times {\bm E}$)
called the Bianchi identities
\begin{eqnarray}
\nabla \cdot (\nabla \cdot {\bf M}) & = & 0 
\nonumber
\\
\nabla\cdot (\nabla \cdot {\bf F}) & = & 0
\nonumber
\end{eqnarray}
These identities are related to J. A. Wheeler's classic statement that
``the boundary of a boundary is zero''.  But, does the staggered grid in
Figure 3 automatically satisfy these identities to machine accuracy? A
quick analysis of $\nabla \cdot {\bf F}$ shows that, in fact, it does.
$\nabla \cdot {\bf F}$ is a vector that is defined on cell edges at
whole time steps and on cell corners at half time steps, and it involves
tensor components that are two half-steps away from scalar points (whole
time-step cell corners). Taking the divergence of this vector causes
like components to cancel, so that
\begin{eqnarray}
\nabla \cdot (\nabla \cdot {\bf F}) & = & O(\epsilon_r)
\nonumber
\end{eqnarray}
holds in this differencing scheme.

Because of the zero on the right hand side of equation
(\ref{ed_cov_magnetic}), that equation itself also is often described
as a Bianchi identity
\begin{eqnarray}
\label{dF_bi}
d {\bf F} & = & 0
\end{eqnarray}
where $d {\bf F}$ is the differential of the tensor ${\bf F}$, which in
component form is given by
\begin{eqnarray}
(d {\rm F})_{\alpha \beta \gamma} & \equiv & {\rm F}_{[ \alpha \beta \, , \, \gamma]} 
~ = ~ {\rm F}_{\alpha \beta \, , \, \gamma} ~ + ~ {\rm F}_{\beta \gamma \, , \, \alpha} ~ + ~ 
{\rm F}_{\gamma \alpha \, , \, \beta} 
\nonumber
\end{eqnarray}
where the comma denotes ordinary differentiation (${\rm F}_{\alpha \beta
\, , \, \gamma} \equiv \partial {\rm F}_{\alpha \beta}  / \partial
x^{\gamma}$) and the brackets denote permuted summation.\footnote{In
this paper Greek indices range from 0 to 3, Roman indices $i, \, j, \,
k, \, ...$ range from 1 to 3, and Roman indices $a, \, b, \, c, \, ...$
will be used to denote a set of three integers with one of the spatial
indices missing ({\it i.e.}, one of the sets [0, 1, 2], [0, 1, 3], or
[0, 2, 3]).} This allows the Faraday tensor to be derived from a vector
four-potential 
\begin{eqnarray}
{\bf F} & = & d {\bf A}
\nonumber
\end{eqnarray}
or
\begin{eqnarray}
{\rm F}_{\alpha \beta} & = & {\rm A}_{\alpha \, , \, \beta}  ~  -  ~  {\rm A}_{\beta \, , \, \alpha}
\nonumber
\end{eqnarray}
This is the covariant form for equations (\ref{initial_b}) and
(\ref{ed_gauge}).  Does the staggered grid automatically enforce $d d
{\bf A} = 0$ also?  Yes.  We have already shown this to be the case for
the magnetic part (equation \ref{divb_discrete}).  For the electric part
of $d d {\bf A} = 0$ we have
\begin{eqnarray}
\nabla \times {\bm E} & = & - \nabla \times \nabla \phi ~ - ~ \nabla \times \frac{1}{c} {\dot {\bm A}}
\nonumber 
\\ 
& = & \frac{1}{c} \dot{\bm B} ~ + ~ O(\epsilon_r)
\label{dda_electric}
\end{eqnarray}
the second term is ${\dot {\bm B}}/c$ to order $\epsilon_r$, and the
first term is zero to the same order because of the staggered grid (see
the $~t =~ ^{-\frac{1}{2}}t$ panel in Figure \ref{fig2}).  So equation
(\ref{dda_electric}) is just the same Faraday's law that we are solving to
machine accuracy in the spatial part of equation (\ref{ed_cov_magnetic})
(or equation \ref{mhd_evolution}).

To summarize, then, the staggered grid naturally satisfies the Bianchi
(and vector) identities in space and time
\begin{eqnarray}
\label{ed_bi1}
{{\rm F}^{\alpha \beta}}_{,\, \beta \alpha} & = & 0 
\\
({\rm A}_{\alpha \, , \, \beta} ~ - ~ {\rm A}_{\beta \, , \, \alpha})_{,\, \gamma} ~ + ~ 
({\rm A}_{\gamma \, , \, \alpha} ~ - ~ {\rm A}_{\alpha \, , \, \gamma})_{,\, \beta} ~ + ~ 
({\rm A}_{\beta \, , \, \gamma}~ - ~ {\rm A}_{\gamma \, , \, \beta})_{,\, \alpha} & = & 0
\label{ed_bi2}
\end{eqnarray}
to machine accuracy for any antisymmetric tensor ${\bf F}$ and for any
four-vector ${\bf A}$.  Here we use the Einstein summation convention,
where a repeated index indicates summation over the four coordinates
(${{\rm F}^{\alpha \beta}}_{\, , \beta} \equiv \Sigma^{\beta = 3}_{\beta
= 0} \, \partial {{\rm F}^{\alpha \beta}} / \partial x^{\beta}$).
Note that equation (\ref{ed_bi1}) uses the raised form of ${\bf F}$,
but in flat space this involves only multiplying by $\pm 1$ with the
Minkowski metric.  As a result of this cancellation, when the spatial
parts of equations (\ref{ed_cov_magnetic}) and (\ref{ed_cov_electric})
are integrated forward in time
\begin{eqnarray}
{\bf P} \cdot (\nabla \cdot {\bf F}) & = & 4 \pi {\bf P} \cdot {\bf J} ~~~~~~~~~~~~~~~~~~~~~~~ 
{\bf P} \cdot (\nabla \cdot {\bf M}) ~ = ~ 0
\nonumber
\end{eqnarray}
(where the spatial projection tensor ${\rm P}^{\alpha \beta} = {\rm n}^{\alpha}
{\rm n}^{\beta} + {\rm g}^{\alpha \beta}$ is orthogonal to ${\bf n}$), the time parts
(the constraints) are {\em automatically satisfied to machine accuracy}
\begin{eqnarray}
{\bf n} \cdot (\nabla \cdot {\bf F}) & = & 4 \pi {\bf n} \cdot {\bf J}   ~~~~~~~~~~~~~~~~~~~~~~~ {\bf n} \cdot 
(\nabla \cdot {\bf M}) ~ = ~ 0
\nonumber
\end{eqnarray}
with no additional computation required.  {\em A staggered grid,
therefore, has ``deep geometric significance''}, because it naturally
satisfies the Bianchi identities.

\section{CT for Electrodynamics in Curvilinear and Curved Spacetime}

CT also works in curvilinear coordinates and in curved spacetime, but
the method requires iteration.  Again, Maxwell's equations are
\begin{eqnarray}
\nabla \cdot {\bf F} & = & 4 \pi {\bf J} ~~~~~~~~~~~~~~~~~~~~~~~ d {\bf F} ~ = ~ 0
\nonumber
\end{eqnarray}
but the gradient operator is now the covariant derivative rather than
the ordinary derivative
\begin{eqnarray}
\label{maxwell_curved1}
{{\rm F}^{\alpha \beta}}_{;\, \beta} & = & 4 \pi {\rm J}^{\alpha} ~~~~~~~~~~~~~~~~~~~~~~~ 
{\rm F}_{[\alpha \beta ;\, \gamma]} ~ = ~ 0
\end{eqnarray}
where 
\begin{eqnarray}
{{\rm T}^{\alpha \beta}}_{;\, \gamma} & \equiv & {{\rm T}^{\alpha \beta}}_{,\, \gamma} 
~ + ~ {\Gamma^{\alpha}}_{\mu \gamma} \, {\rm T}^{\mu \beta} 
~ + ~ {\Gamma^{\beta}}_{\mu \gamma} \, {\rm T}^{\alpha \mu} 
\nonumber
\\
{\rm T}_{\alpha \beta ;\, \gamma} & \equiv & {\rm T}_{\alpha \beta \, , \, \gamma}
~ - ~ {\Gamma^{\mu}}_{\alpha \gamma} \, {\rm T}_{\mu \beta} 
~ - ~ {\Gamma^{\mu}}_{\beta \gamma} \, {\rm T}_{\alpha \mu} 
\nonumber
\end{eqnarray}
are the covariant derivatives of a general second-rank tensor ${\bf T}$,
\begin{eqnarray}
\label{singly_raised_Gamma}
{\Gamma^{\alpha}}_{\beta \gamma} & \equiv & {\rm g}^{\alpha \mu} {\Gamma_{\mu \beta \gamma}}
\\
{\Gamma}_{\alpha \beta \gamma} & \equiv & \frac{1}{2} \left( 
{\rm g}_{\alpha \beta \, , \, \gamma} + {\rm g}_{\beta \alpha \, , \, \gamma} - {\rm g}_{\beta \gamma \, , \, \alpha} \right)
\label{non_raised_Gamma}
\end{eqnarray}
are different forms of the connection coefficients, and ${\rm g}^{\alpha \beta}$
is the inverse of the metric ${\rm g}_{\alpha \beta}$.  However, because the
Faraday tensor is antisymmetric (as is the Maxwell tensor), equations
(\ref{maxwell_curved1}) reduce to
\begin{eqnarray}
\label{maxwell_curved2}
{{\rm F'~}^{\alpha \beta}}_{,\, \beta} & = & 4 \pi {\rm J'~}^{\alpha} 
~~~~~~~~~~~~~~~~~~~~~~~ {\rm F}_{[\alpha \beta \, , \, \gamma]} ~ = ~ 0
\end{eqnarray}
where 
\begin{eqnarray}
{{\rm F'~}^{\alpha \beta}} \equiv {{\rm F}^{\alpha \beta}} \sqrt{-g}
~~~~~~~~~~~~~~~~~~~~~~~ 
{{\rm J'~}^{\alpha}} \equiv {\rm J}^{\alpha} \sqrt{-g}
\nonumber
\end{eqnarray}
and $\sqrt{-g}$ is the volume element (square root of the negative
metric determinant).  Equations (\ref{maxwell_curved2}) involve only
simple differences and known values for the metric.  However, unless
${\bf g}$ is a diagonal, the raised version of the Faraday tensor ${\rm
F}^{\alpha \beta}$ involves the sum of several lowered version components
at different grid locations
\begin{eqnarray}
{{\rm F}^{\alpha \beta}} & \equiv & {\rm g}^{\alpha \mu} \, {\rm g}^{\beta \nu} \, {{\rm F}_{\mu \nu}}
\nonumber
\end{eqnarray}
Therefore, the time update of the fundamental variables ${{\rm F}_{\mu
\nu}}$ will necessarily be implicit, and therefore iterative, as it
involves sums over time as well as space.

It is important to realize, however, that even in curvilinear coordinates
and curved spacetime, the Bianchi identities
\begin{eqnarray}
\label{curved_bi}
{{\rm F'~}^{\alpha \beta}}_{,\, \beta \alpha} & = & 0 
\end{eqnarray}
will be satisfied to machine accuracy, because ${{\rm F'~}^{\alpha
\beta}}$ is constructed {\em before} it is differenced and because
that differencing is done in precisely the same manner as in equation
(\ref{ed_bi1}).  It does not matter that ${{\rm F'~}^{\alpha \beta}}$
involves the sum of many tensor components and products of metric
components.  It only matters that the $\beta$ and $\alpha$ derivatives
commute.

\section{A General Finite Difference Prescription for CT for Tensor Field 
Evolution Problems}
\label{general_scheme}

\begin{figure}
\begin{center}
\centerline{\psfig{figure=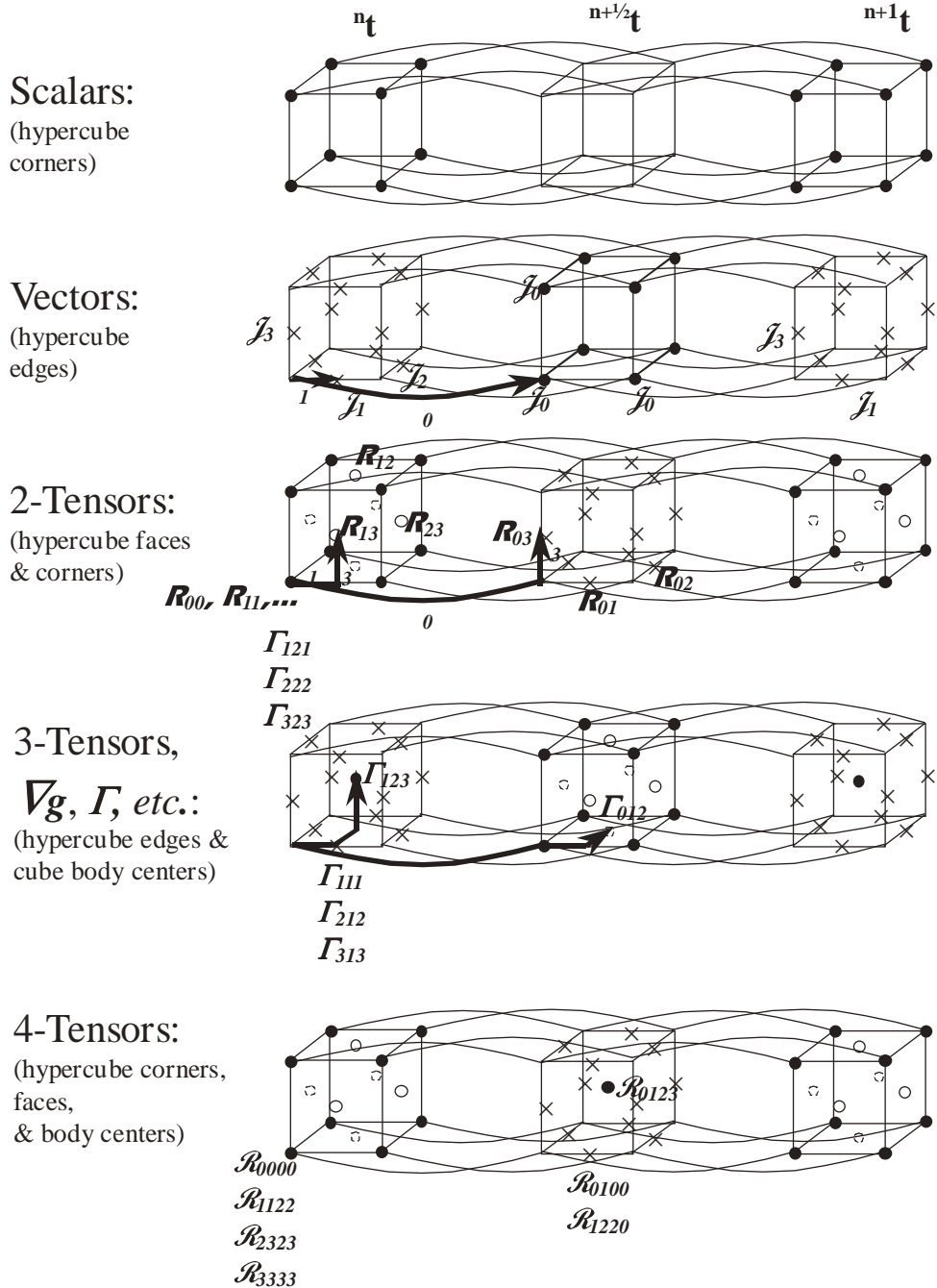,height=7.0in}}
\end{center}
\caption{Graphical depiction of the four-dimensional staggered grid scheme
for different tensor quantities.  (See Section \ref{general_scheme}
for a full description.)  The first panel shows filled circles where
scalars are located (hypercube corners.)  Heavy arrows in the second
panel show how to locate the following vector quantities:  ${\rm J}_{0}$
(shift one-half step in the $0$ dimension); ${\rm J}_{1}$ (shift one-half
step in the $1$ dimension).  The third panel shows how to locate ${\rm
R}_{0 3}$ by shifting one-half step in the $0$ and $3$ dimensions, and
${\rm R}_{1 3}$ by shifting in the $1$ and $3$ dimensions. Remaining
panels demonstrate third and fourth ranked tensors. Note the hypercube
body-center location of ${\rm R}_{\, 0 \, 1 \, 2 \, 3}$.
\label{fig4}}
\end{figure}

Figure \ref{fig3} suggests the following geometric prescription for a
staggered grid when solving covariant tensor field evolution problems.
This prescription is depicted schematically in Figure \ref{fig4}, and
appears to work for tensors up to at least rank five.  The basic rules are

\begin{enumerate}
\item{Extend and stagger the grid in time as well as in space.  The time
extension need not be very deep --- only enough cells to compute the
tensor components, derivatives, {\it etc.}  to the order of the method.
In this paper we use a second order differencing scheme, so we need only
one additional half, and one full, time slice.} 
\item{Treat time otherwise like a spatial coordinate.  That is, use the
same differencing scheme in time as used in space so that temporal and
spatial differences commute.}
\item{Starting at corner nodes on the four-dimensional hypercube cell,
define the following quantities in the manner described below.  Simply
put, a tensor component is located one-half step away from hypercube
cell corners in each dimension specified by that component's indices;
two repeated indices are equivalent to no shift at all.
\begin{enumerate}
\item{Scalars: located on 4-cube cell corners (3-cube corners at whole
time steps)}
\item{Vectors: located on 4-cube edges, shifted one-half cell step in
the direction specified by that component.  That is,
\begin{enumerate}
\item{${\rm J}^0$ on 3-cube cell corners at half time steps}
\item{${\rm J}^1$ centered on 3-cube $x$ edges at whole time steps}
\item{${\rm J}^2$ centered on $y$ edges}
\item{${\rm J}^3$ centered on $z$ edges}
\end{enumerate}
}
\item{One-forms: defined like vectors}
\item{Second-ranked tensors of any type
\begin{enumerate}
\item{${\rm R}_{\alpha \alpha}$: defined like scalars, on 3-cube cell corners
at whole time steps}
\item{${\rm R}_{0 \, i}$:  centered on 3-cube $i$-edges at {\em half} time steps}
\item{${\rm R}_{i j}$:  centered on $ij$ faces at whole time steps}
\end{enumerate}
}
\item{Third-ranked tensors and connection coefficients
\begin{enumerate}
\item{$\Gamma_{\alpha \alpha \beta}$, $\Gamma_{\alpha \beta \alpha}$,
$\Gamma_{\beta \alpha \alpha}$: defined like ${\rm J}^\beta$}
\item{$\Gamma_{0 \, i j}$, $\Gamma_{i \, 0 \, j}$, $\Gamma_{i j \, 0}$:
centered on 3-cube $ij$-faces at half time steps}
\item{$\Gamma_{i j \, k}$:  3-cube-centered at whole time steps}
\end{enumerate}
}
\item{Fourth-ranked tensors
\begin{enumerate}
\item{${\rm R}_{\, \alpha \alpha \beta \gamma \,}$:  defined like
${\rm R}_{\beta \gamma}$}
\item{${\rm R}_{\, \alpha \beta \gamma \delta}$ ($\alpha \neq \beta
\neq \gamma \neq \delta$): located at 4-cube body centers}
\end{enumerate}
}
\item{Bianchi identities:  defined like third-ranked tensors (at least
one index must be repeated).  Examples are ${\rm R}_{\, 0 0 2 3 ; 2}$
(same as ${\rm J}^3$) and ${\rm R}_{\, 0 1 2 3 ; 2}$, (centered on
the 1-3 face at half time steps).}
\end{enumerate}
}
\end{enumerate}

This differencing scheme has a number of properties that make it
look like a finite implementation of differential calculus.  First,
the differential operator, which creates a tensor of one higher order
({\it e.g.}, equation \ref{dF_bi}), naturally places the new tensor on
the proper grid if the differencing is centered.  This is also true
of the generation of ${\Gamma_{\alpha \, \beta \, \gamma}}$ from the
metric field ${\rm g}_{\alpha \beta}$.  Second, one of the most fundamental 
properties of a spacetime, the covariant derivative of the metric 
(${\rm g}_{\alpha \, \beta \, ; \, \gamma}$) vanishes, is naturally  
satisfied to machine accuracy because of this property.  Third, the 
contraction of a mixed tensor is trivial:  for each staggered component 
of the contracted tensor the four components of the parent tensor that 
are needed for the sum are already located at the same grid point as the 
contracted component.  No additional averaging is needed.  Fourth, 
as shown below, in a local Riemann normal coordinate system, the Bianchi 
identities are satisfied to machine accuracy.  All terms cancel exactly, 
so the constraints are propagated exactly as well. 

Certain special tensors also have interesting properties.  The Kronecker
delta ${\delta^{\alpha}}_{\beta}$, for example, has non-zero elements
(unity) only at hypercube cell corners.  This is also true of other
identity tensors (${\delta^{\alpha \beta}}_{\lambda \, \mu}$,
${\delta^{\alpha \beta \gamma}}_{\lambda \, \mu \, \nu}$, {\it
etc.}), which are $\pm 1$ at cell corners.  The Levi-Civita tensor
$\epsilon_{\alpha \, \beta \, \gamma \, \delta}$ and antisymmetric symbol
$[\alpha \beta \gamma \delta]$ are just the opposite.  They are zero
everywhere except at the hypercube {\em body centers}.  They look much
like the identity tensors, but on a grid that is shifted one-half step
in each dimension.  Furthermore, as the Levi-Civita tensor expresses
the volume element
\begin{eqnarray}
\epsilon_{\alpha \, \beta \, \gamma \, \delta} & = & 
\sqrt{-g} ~ [\alpha \beta \gamma \delta]
\nonumber
\end{eqnarray}
its placement at the hypercube center creates a natural scheme for
forming volume integrals over those hypercubes.

\section{CT for Numerical Relativity}

\subsection{Statement of the Problem} 
\label{prob_statement}

For reasons that are developed more fully below, we will use mixed
tensors to define the problem of numerical relativity.  As noted above,
such tensors will lend themselves easily to contraction.

The classic problem of general relativity is to solve Einstein's field
equations
\begin{eqnarray}
\label{ein_field}
{{\rm G}^{\alpha}}_{\beta} & = & 8 \pi \, {{\rm T}^{\alpha}}_{\beta}
\end{eqnarray}
($c = G = 1$) for the metric coefficients.  The Einstein curvature tensor
is derived from the Ricci curvature tensor
\begin{eqnarray}
{{\rm G}^{\alpha}}_{\beta} & \equiv {{\rm R}^{\alpha}}_{\beta} ~ - ~\frac{1}{2} 
{\delta^{\alpha}}_{\beta} ~ {R}
\end{eqnarray}
with
\begin{eqnarray}
{R} & \equiv & {{\rm R}^{\alpha}}_{\alpha}
\end{eqnarray}
being the Ricci curvature scalar.  The Ricci tensor is the contraction
of the Riemann tensor on the first and third indices
\begin{eqnarray}
{{\rm R}^{\alpha}}_{\beta} & \equiv & {{\rm R}^{\mu \, \alpha}}_{\mu \, \beta}
\end{eqnarray}
The Riemann tensor is the full statement of curvature of the spacetime.
In its mixed form it is given by
\begin{eqnarray}
\label{mixed_riemann}
{{\rm R}^{\alpha \beta}}_{\gamma \delta} & = & 
{\Gamma^{\alpha \beta}}_{\delta ,~ \gamma} ~ - ~
{\Gamma^{\alpha \beta}}_{\gamma ,~ \delta} ~ + ~
{\Gamma^{\alpha \mu}}_{\delta} \, {\Gamma^{\beta}}_{\mu \gamma} ~ - ~
{\Gamma^{\alpha \mu}}_{\gamma} \, {\Gamma^{\beta}}_{\mu \delta}
\end{eqnarray}
for a coordinate basis.  The doubly-raised connection coefficients are
given by
\begin{eqnarray}
\label{doubly_raised_Gamma}
{\Gamma^{\alpha \beta}}_{\gamma} & \equiv & 
{\rm g}^{\beta \nu} \, {\Gamma^{\alpha}}_{\nu \gamma}
\end{eqnarray}
We assume that the metric ${\rm g}_{\alpha \beta}$ has a unique inverse
${\rm g}^{\alpha \beta}$ such that
\begin{eqnarray}
{\rm g}^{\alpha \mu} \, {\rm g}_{\mu \beta} & = & {\delta^{\alpha}}_{\beta}
\end{eqnarray}
In this paper we will treat only the vacuum problem (the source of
stress-energy ${{\rm T}^{\alpha}}_{\beta} = 0$) so that equation
(\ref{ein_field}) becomes
\begin{eqnarray}
{{\rm R}^{\alpha}}_{\beta} & = & {{\rm G}^{\alpha}}_{\beta} ~ = ~ 0
\end{eqnarray}
The Riemann tensor ${{\rm R}^{\alpha \beta}}_{\gamma \, \delta}$
possesses several symmetries, including algebraic antisymmetry on $\alpha$
and $\beta$ (and on $\gamma$ and $\delta$) and differential symmetries
(Bianchi identities)
\begin{eqnarray}
\label{full_bi}
{{\rm R}^{\alpha \beta}}_{[\gamma  \, \delta   ;\, \epsilon]} & \equiv & 
{{\rm R}^{\alpha \beta}}_{\gamma   \, \delta   ;\, \epsilon} ~ + ~ 
{{\rm R}^{\alpha \beta}}_{\epsilon \, \gamma   ;\, \delta} ~ + ~
{{\rm R}^{\alpha \beta}}_{\delta   \, \epsilon ;\, \gamma} ~ = ~ 0
\end{eqnarray}
The reader will note that the mixed Riemann tensor is missing one
additional symmetry that is possessed by the covariant version:
${{\rm R}}_{\alpha \beta \gamma \delta} = {{\rm R}}_{\gamma \delta
\alpha \beta}$.  (A raised index cannot be swapped with a lower one.)
When contracted on the first and third indices, the Bianchi identities
become simply
\begin{eqnarray}
\label{rn_contracted_bi}
{{\rm R}^{\alpha}}_{\beta \, ; \, \alpha} ~ - ~ 
\frac{1}{2} {{R}}_{   ~  , \, \beta} & = & 
{{\rm G}^{\alpha}}_{\beta \, ; \, \alpha} ~ = ~ 0
\end{eqnarray}
{\it i.e.}, the divergence free condition on the Einstein tensor.
These four conditions are responsible for propagating the four constraints
\begin{eqnarray}
{{\rm G}^{ 0 }}_{ \beta } & = & 0
\end{eqnarray}
if the other equations are satisfied
\begin{eqnarray}
{{\rm G}^{ i }}_{ \beta } & = & 0
\end{eqnarray}

\subsection{CT in Riemann Normal Coordinates}
\label{ct_in_rn}

At any point $P$ in spacetime one can construct many transformations
${{\rm L}^{\alpha}}_{\hat{\beta}}$ to locally Lorentz systems such that
\begin{eqnarray}
{\rm g}_{\hat{\alpha} \hat{\beta}} & = &
{{\rm L}^{\mu}}_{\hat{\alpha}} \, {{\rm L}^{\nu}}_{\hat{\beta}} \, {\rm g}_{\mu \nu} 
~ = ~ \eta_{\hat{\alpha} \hat{\beta}}
\end{eqnarray}
in a neighborhood of that point.  However, only one of those systems
--- the Riemann normal system --- also has vanishing {\em gradients}
of the metric and, therefore, vanishing connection coefficients in that
same neighborhood
\begin{eqnarray}
{\rm g}_{\hat{\alpha} \, \hat{\beta} \, , \, \hat{\gamma}} & = & 0
\nonumber
\\
\Gamma_{\hat{\alpha} \, \hat{\beta} \, \hat{\gamma}} & = & 0
\nonumber
\end{eqnarray}
In this coordinate system, in the neighborhood of $P$, covariant
derivatives become ordinary derivatives and the Riemann tensor and its
Bianchi identities become
\begin{eqnarray}
{{\rm R}^{\hat{\alpha} \hat{\beta}}}_{\hat{\gamma} \hat{\delta}} & = & 
{\Gamma^{\hat{\alpha} \hat{\beta}}}_{\hat{\delta} \, , \, \hat{\gamma}} ~ - ~ 
{\Gamma^{\hat{\alpha} \hat{\beta}}}_{\hat{\gamma} \, , \, \hat{\delta}}
\\
{{\rm R}^{\hat{\alpha} \hat{\beta}}}_{[ \hat{\gamma}   \,      \hat{\delta}   \, ; \, \hat{\epsilon}]} & = & 
{\Gamma^{\hat{\alpha} \hat{\beta}}}_{    \hat{\delta}   \, , \, \hat{\gamma}   \, \hat{\epsilon}} ~ - ~ 
{\Gamma^{\hat{\alpha} \hat{\beta}}}_{    \hat{\gamma}   \, , \, \hat{\delta}   \, \hat{\epsilon}} ~ + ~ 
{\Gamma^{\hat{\alpha} \hat{\beta}}}_{    \hat{\gamma}   \, , \, \hat{\epsilon} \, \hat{\delta}} ~ - ~ 
{\Gamma^{\hat{\alpha} \hat{\beta}}}_{    \hat{\epsilon} \, , \, \hat{\gamma}   \, \hat{\delta}} ~ + ~ 
{\Gamma^{\hat{\alpha} \hat{\beta}}}_{    \hat{\epsilon} \, , \, \hat{\delta}   \, \hat{\gamma}} ~ - ~ 
{\Gamma^{\hat{\alpha} \hat{\beta}}}_{    \hat{\delta}   \, , \, \hat{\epsilon} \, \hat{\gamma}} ~ = ~ 0
~~~
\label{rn_bi}
\end{eqnarray}
In the proposed staggered grid scheme in the previous section, each of
the terms of ${{\rm R}^{\hat{\alpha} \hat{\beta}}}_{[ \hat{\gamma} \,
\hat{\delta} ; \, \hat{\epsilon}]}$ would be evaluated at the same grid
point, because they each have the same five indices.  So the sum can
be accomplished without additional averaging from other grid points.
We further note that each term has a duplicate with the opposite
sign, differing only in the order of the derivatives ({\it e.g.},
${\Gamma^{\hat{\alpha} \hat{\beta}}}_{ \hat{\delta} \, , \, \hat{\gamma}
\, \hat{\epsilon}} ~ - ~ {\Gamma^{\hat{\alpha} \hat{\beta}}}_{\hat{\delta}
\, , \,  \hat{\epsilon} \, \hat{\gamma}}$).  We can therefore draw the
following conclusion:  {\em If a numerical scheme is constructed such that
derivatives commute (both space-space and space-time), then in Riemann
normal coordinates the equation for the Bianchi identities (\ref{rn_bi})
will be satisfied to machine accuracy, resulting in the propagation of
the constraints to machine accuracy.}  We note that the scheme proposed
in Section \ref{general_scheme} possesses the required properties.

How does satisfying the Bianchi identities propagate the constraints
numerically?  This is easy to show in Riemann normal coordinates.
For the vacuum problem, the constraints are given by
\begin{eqnarray}
{{\rm R}^{\hat{0}}}_{\hat{\beta}} & = & 0
\end{eqnarray}
and we seek a scheme in which the constraints propagate to machine
accuracy
\begin{eqnarray}
{{\rm R}^{\hat{0}}}_{\hat{\beta} \, , \, \hat{0}} & = & O(\epsilon_r)
\end{eqnarray}
But satisfying the Bianchi identities (equation \ref{rn_bi}) will mean
that the contracted identities are also satisfied to machine accuracy
\begin{eqnarray}
{{\rm R}^{\hat{\alpha}}}_{\hat{\beta} \, , \, \hat{\alpha}} & = & O(\epsilon_r)
\end{eqnarray}
or
\begin{eqnarray}
{{\rm R}^{\hat{0}}}_{\hat{\beta} \, , \, \hat{0}} & = & -{{\rm R}^{\hat{i}}}_{\hat{\beta} \, , \, \hat{i}}
~ + ~ O(\epsilon_r)
\end{eqnarray}

All that needs to be shown is that ${{\rm R}^{\hat{i}}}_{\hat{\beta} \,
, \, \hat{i}} ~ = ~ O(\epsilon_r)$ also.  For the three momentum
constraints ($\hat{\beta} = \hat{j}$) this is straightforward,
because ${{\rm R}^{\hat{i}}}_{\hat{j}} ~ = ~ 0$ are the field equations
being computed, so ${{\rm R}^{\hat{i}}}_{\hat{j}}$ is $O(\epsilon_r)$ by 
definition.\footnote{This statement requires a little clarification. 
Of course, the {\em solution} of ${{\rm R}^{\hat{i}}}_{\hat{j}} ~ = ~ 0$ is 
accurate to only $O(\epsilon_{tr})$.  However, if we use this 
truncation-accurate solution to re-compute ${{\rm R}^{\hat{i}}}_{\hat{j}}$, 
{\em using exactly the same mathematical definition of terms that we used in 
the evolution equation}, then that re-computed ${{\rm R}^{\hat{i}}}_{\hat{j}}$ 
will be zero to machine accuracy.  (It will not be so only if we use a different 
differencing scheme than the one used in the original evolution equation.)
As a simple example, consider a line of code that computes $y ~ = ~ a \, x \, + \, b$.
Then, if we later compute the function $f ~ = ~ y \, - \, a \, x \, - \, b$, 
by definition, $f$ will be zero to machine accuracy.}
So the spatial gradients ${{\rm R}^{\hat{i}}}_{\hat{j} \, , \, \hat{i}}$
will be $O(\epsilon_r)$, thereby propagating ${{\rm R}^{\hat{0}}}_{\hat{j}}$
to machine accuracy.

For the Hamiltonian constraint ($\hat{\beta} = 0$) we require
satisfaction of the momentum constraints on the hypersurface, {\it i.e.},
${{\rm R}^{\hat{i}}}_{\hat{0} \, , \, \hat{i}} = O(\epsilon_r)$.  Therefore,
as long as the momentum constraints propagate to machine accuracy on each
hypersurface, which we have shown above to be the case, the Hamiltonian
constraint will propagate also.  Note, however, if the gradient of the
momentum constraints has a constant bias, then the Hamiltonian constraint
will grow. On the other hand, if the divergence fluctuates randomly,
the the Hamiltonian constraint will fluctuate only randomly as well.

\subsection{CT in a General Coordinate Basis}

Because a Riemann normal coordinate system is local only and cannot be
used to cover the entire spacetime, we are forced to deal with non-zero
connection coefficients.  But we still will attempt to propagate the
constraints in the same manner --- with a scheme that enforces the Bianchi
identities by employing temporal and spatial differences that commute.

\subsubsection{Propagation of the Constraints}

It is important to note that the Bianchi identities are never {\em
explicitly} calculated in the CT method.  Instead, we develop a scheme
in which
\begin{eqnarray}
{{\rm R}^{0}}_{\beta \, , \, 0} & = & 0
\nonumber
\end{eqnarray}
is satisfied to at least truncation accuracy for all time.  (We will find
that machine accuracy may be an unattainable goal.)  However, in a general
coordinate system, the update of ${{\rm R}^{0}}_{\beta}$ will depend on an
``advection'' of curvature from surrounding cells, due to the covariant
derivative connection terms.  So, we equivalently seek a scheme in which
\begin{eqnarray}
{{\rm R}^{0}}_{\beta \, ; \, 0} & = & 0
\end{eqnarray}
to at least truncation accuracy.  If we set up a grid in which 
\begin{eqnarray}
{{\rm R}^{\alpha \beta}}_{ [ \gamma \, \delta \, ; \, \epsilon ] } & = & O(\epsilon_{tr})
\end{eqnarray}
then we can proceed in much the same manner as in Riemann normal
coordinates, but with covariant derivatives.  If the Bianchi identities
are satisfied to truncation accuracy, then their contracted form also
will hold
\begin{eqnarray}
{{\rm R}^{0}}_{0 \, ; \, 0} & = -{{\rm R}^{j}}_{0 \, ; \, j} ~ + ~ O(\epsilon_{tr})
\nonumber
\\
{{\rm R}^{0}}_{i \, ; \, 0} & = -{{\rm R}^{j}}_{i \, ; \, j} ~ + ~ O(\epsilon_{tr})
\nonumber
\end{eqnarray}
So, if the momentum constraints are satisfied on each hypersurface,
the Hamiltonian constraint will propagate, albeit along geodesics, not
coordinate lines.  And the momentum constraints will propagate if the
field equations are satisfied to at least truncation accuracy.

\subsubsection{Cancellation of the $\partial^2 \Gamma$ Terms}

In order to demonstrate cancellation of terms in the Bianchi identities,
we choose the following form for the mixed Riemann tensor
\begin{eqnarray}
{{\rm R}^{\alpha \beta}}_{\gamma \, \delta} & = & \frac{1}{2} 
\left( {\Gamma^{ \alpha \beta }}_{\delta \, , \, \gamma} ~ - ~ 
{\Gamma^{ \alpha \beta }}_{ \gamma \, , \, \delta } ~ - ~ 
{\Gamma^{ \beta \alpha }}_{ \delta \, , \, \gamma } ~ + ~ 
{\Gamma^{ \beta \alpha }}_{ \gamma \, , \, \delta } \right.
\nonumber
\\
& & ~ + ~ \left.
{\Gamma^{ \alpha \mu }}_{ \delta } \, {\Gamma^{ \beta }}_{ \mu \, \gamma } ~ - ~ 
{\Gamma^{ \alpha \mu }}_{ \gamma } \, {\Gamma^{ \beta }}_{ \mu \, \delta } ~ + ~ 
{\Gamma^{ \alpha }}_{ \mu \delta } \, {\Gamma^{ \beta \, \mu }}_{ \gamma } ~ - ~ 
{\Gamma^{ \alpha }}_{ \mu \gamma } \, {\Gamma^{ \beta \, \mu }}_{ \delta } \right)
\label{new_mixed_riemann}
\end{eqnarray}
with the singly-raised connection coefficient {\em re}-computed from
the doubly-raised ones and the gradient of the inverse metric
\begin{eqnarray}
\label{alt_raised_Gamma}
{\Gamma^{\beta}}_{\mu \, \gamma} & = & \frac{{\rm g}_{\mu \, \nu}}{2} 
\left( {\Gamma^{ \beta \nu }}_{ \gamma } ~ - ~ 
{\Gamma^{ \nu \beta }}_{ \gamma } ~ - ~ 
{{\rm g}^{\beta \nu}}_{ \, , \, \gamma} \right)
\end{eqnarray}
This form has the following properties
\begin{enumerate}
\item{${{\rm R}^{\alpha \beta}}_{\gamma \, \delta}$ possesses explicitly all of 
the algebraic symmetries discussed in Section \ref{prob_statement}.}
\item{When the Bianchi identities are formed\footnote{Note that,
because of the antisymmetry in the last two indices of ${{\rm R}^{\alpha
\beta}}_{\gamma \, \delta}$ and the symmetry in the last two indices of
${\Gamma^{\alpha}}_{\beta \, \gamma}$, the connection terms involving
$\gamma$ and $\delta$ will cancel, just as they did for the antisymmetric
Faraday tensor Bianchi identities in equation (\ref{maxwell_curved2}).}
\begin{eqnarray}
{{\rm R}^{\alpha \beta}}_{ \gamma \, \delta \, , \epsilon } ~ + ~ 
{\Gamma^{ \alpha }}_{ \mu \, \epsilon } \, {{\rm R}^{\mu \beta}}_{ \gamma \, \delta } ~ + ~ 
{\Gamma^{ \beta }}_{ \mu \, \epsilon } \, {{\rm R}^{\alpha \mu}}_{ \gamma \, \delta } ~ + ~ 
& &
\nonumber
\\
{{\rm R}^{\alpha \beta}}_{ \epsilon \, \gamma \, , \delta } ~ + ~ 
{\Gamma^{ \alpha }}_{ \mu \, \delta } \, {{\rm R}^{\mu \beta}}_{ \epsilon \, \gamma } ~ + ~ 
{\Gamma^{ \beta }}_{ \mu \, \delta } \, {{\rm R}^{\alpha \mu}}_{ \epsilon \, \gamma } ~ + ~ 
& &
\nonumber
\\
{{\rm R}^{\alpha \beta}}_{ \delta \, \epsilon \, , \gamma } ~ + ~ 
{\Gamma^{ \alpha }}_{ \mu \, \gamma } \, {{\rm R}^{\mu \beta}}_{ \delta \, \epsilon } ~ + ~ 
{\Gamma^{ \beta }}_{ \mu \, \gamma } \, {{\rm R}^{\alpha \mu}}_{ \delta \, \epsilon } 
~~~~~~ & = & 0
\label{cb_full_bi}
\end{eqnarray}
and equation (\ref{alt_raised_Gamma}) is inserted, the 
${\Gamma^{\alpha \beta}}_{\gamma \, , \, \delta \, \epsilon}$ terms will cancel 
{\em to machine accuracy}, as described before in section \ref{ct_in_rn}.
}
\end{enumerate}

\subsubsection{Cancellation of the $\Gamma \, \partial \Gamma$ Terms}

In equation (\ref{cb_full_bi}) the 
${\Gamma^{\alpha}}_{\mu \, \epsilon} \, {\Gamma^{\mu \beta}}_{\delta \, , \, \gamma}$ 
terms will cancel explicitly {\em algebraically}.  Will they cancel numerically also?  
The answer is yes, if we apply the following numerical procedures
\begin{enumerate}
\item{Use linear interpolation (averaging) to determine quantities at intermediate 
grid points}
\item{When forming a product, such as ${\rm g}_{\mu \, \nu} \, {\Gamma^{\beta
\nu}}_{\gamma}$, first average the factors to the grid point in question,
then form the products and finally the sum.  Do not form the products
on different grid points and then average.}
\end{enumerate}
Consider, for example, the two following Bianchi identity terms
\begin{eqnarray}
\label{bi_terms}
{\Gamma^{\alpha \mu}}_{ \delta \, , \, \epsilon } \, {\Gamma^{\beta}}_{ \mu \, \gamma }
~ - ~ 
{\Gamma^{\beta}}_{ \mu \, \gamma } \, {\Gamma^{\alpha \mu}}_{ \delta \, , \, \epsilon } 
\end{eqnarray}
Algebraically, of course, the two terms cancel.  However, in a staggered
grid scheme they are not computed in the same manner.  The first term
comes from {\em differencing} a $\Gamma \, \Gamma$ term in the first term
of equation (\ref{cb_full_bi}), while the second comes from the {\em
connection} of one of the $\partial \Gamma$ terms in the last term of
that same equation.  One is the difference of an average, while the other
is the average of differences.  But, with linear averaging, we see that,
under these conditions, the chain rule is satisfied to machine accuracy
\begin{eqnarray}
_{_0} \left[ {\Gamma^{\alpha \mu}}_{\delta} \, {\Gamma^{\beta}}_{\mu \, \gamma} 
\right]_{ , \, \epsilon} & = & 
\frac{1}{\Delta x^{\epsilon}}
\left\{ ~~_{_{\frac{1}{2}}} \left[ {\Gamma^{\alpha \mu}}_{\delta} \, {\Gamma^{\beta}}_{\mu \, \gamma} 
\right] ~ - 
~~_{_{-\frac{1}{2}}} \left[ {\Gamma^{\alpha \mu}}_{\delta} \, {\Gamma^{\beta}}_{\mu \, \gamma} 
\right] \, \right\} 
\nonumber
\\
& = & 
\frac{1}{\Delta x^{\epsilon}}
\left\{ \left[ ~_{\frac{1}{2}} {\Gamma^{\alpha \mu}}_{\delta} ~ - 
~_{-\frac{1}{2}} {\Gamma^{\alpha \mu}}_{\delta} \right] ~ 
\left[ {\frac{1}{2}} \left( ~_{\frac{1}{2}} {\Gamma^{\beta}}_{\mu \, \gamma} ~ + 
~_{-\frac{1}{2}} {\Gamma^{\beta}}_{\mu \, \gamma} 
\right) \right] \right.
\nonumber
\\
& & ~~~~ \left. + ~ \left[ \frac{1}{2} \left( ~_{\frac{1}{2}} {\Gamma^{\alpha \mu}}_{\delta} ~ + ~
~_{-\frac{1}{2}} {\Gamma^{\alpha \mu}}_{\delta} \right) \right] 
\left[ ~_{\frac{1}{2}} {\Gamma^{\beta}}_{\mu \, \gamma} ~ - 
~_{-\frac{1}{2}} {\Gamma^{\beta}}_{\mu \, \gamma} \right] \right\}
\nonumber
\\
& = & ~_{_0} \left( {\Gamma^{\alpha \mu}}_{\delta \, , \, \epsilon} \, 
{\Gamma^{\beta}}_{\mu \, \gamma} \right) ~ + ~ 
~_{_0} \left( {\Gamma^{\alpha \mu}}_{\delta} \, 
{\Gamma^{\beta}}_{\mu \, \gamma \, , \epsilon} \right) 
\label{chain_rule}
\end{eqnarray}
where the pre-appended subscript $\frac{1}{2}$ signifies the spatial node position 
at which that quantity is evaluated, {\it e.g.}, $_{\frac{1}{2}} x^{\epsilon} = 
\frac{1}{2} \left( _{_1}x^{\epsilon} + _{_0}x^{\epsilon} \right)$, and the quantities 
$_{\frac{1}{2}} {\Gamma^{\alpha \mu}}_{\delta}$ are themselves averages
\begin{eqnarray}
_{\frac{1}{2}} {\Gamma^{\alpha \mu}}_{\delta} & = & 
\frac{1}{2} \left( _{_1} {\Gamma^{\alpha \mu}}_{\delta} ~ + ~ 
_{_0} {\Gamma^{\alpha \mu}}_{\delta} \right)
\end{eqnarray}
So, the differencing of a $\Gamma \, \Gamma$ term will produce two
averaged $\Gamma \, \partial \Gamma$ terms.  In addition, averaging
factors {\em before} forming products causes the average and difference
operators to commute, so that the second term in (\ref{bi_terms}) becomes
\begin{eqnarray}
-_{_0} \left( {\Gamma^{\beta}}_{\mu \gamma} \, {\Gamma^{\alpha \mu}}_{\delta \, , \, \epsilon} \right)
& = & 
-\frac{1}{2} \left[ 
_{\frac{1}{2}} {\Gamma^{\beta}}_{\mu \gamma} ~ +
_{-\frac{1}{2}} {\Gamma^{\beta}}_{\mu \gamma} \right] \, 
\frac{1}{2} \left[
_{\frac{1}{2}} {\Gamma^{\alpha \mu}}_{\delta \, , \, \epsilon} ~ +
_{-\frac{1}{2}} {\Gamma^{\alpha \mu}}_{\delta \, , \, \epsilon} \right]
\nonumber
\\
& = & -\frac{1}{2} \left[ 
_{\frac{1}{2}} {\Gamma^{\beta}}_{\mu \gamma} ~ +
_{-\frac{1}{2}} {\Gamma^{\beta}}_{\mu \gamma} \right] \, 
\frac{1}{2 \Delta x^{\epsilon}} \left[
\left( _{_1} {\Gamma^{\alpha \mu}}_{\delta} - _{_0} {\Gamma^{\alpha \mu}}_{\delta} \right) ~ +
\left( _{_0} {\Gamma^{\alpha \mu}}_{\delta} - _{_{-1}} {\Gamma^{\alpha \mu}}_{\delta} \right) \right]
\nonumber
\\
& = & -\frac{1}{2} \left[ 
_{\frac{1}{2}} {\Gamma^{\beta}}_{\mu \gamma} ~ + 
_{-\frac{1}{2}} {\Gamma^{\beta}}_{\mu \gamma} \right] \, 
\frac{1}{\Delta x^{\epsilon}} \left[
\frac{1}{2} \left( _{_1} {\Gamma^{\alpha \mu}}_{\delta} + _{_0} {\Gamma^{\alpha \mu}}_{\delta} \right) ~ - ~
\frac{1}{2} \left( _{_0} {\Gamma^{\alpha \mu}}_{\delta} + 
_{_{-1}} {\Gamma^{\alpha \mu}}_{\delta} \right) \right]
\nonumber
\\
& = & -\frac{1}{2} \left[ 
_{\frac{1}{2}} {\Gamma^{\beta}}_{\mu \gamma} ~ +
_{-\frac{1}{2}} {\Gamma^{\beta}}_{\mu \gamma} \right] \, 
\frac{1}{\Delta x^{\epsilon}} \left[
_{\frac{1}{2}} {\Gamma^{\alpha \mu}}_{\delta} ~ - _{-\frac{1}{2}} {\Gamma^{\alpha \mu}}_{\delta} \right]
\nonumber
\\
& = & 
- ~ _{_0} {\Gamma^{\beta}}_{\mu \gamma} ~~ _{_0} {\Gamma^{\alpha \mu}}_{\delta \, , \, \epsilon} 
\nonumber
\end{eqnarray}
which exactly cancels the first term on the right side of equation
(\ref{chain_rule}).  A similar process with another $- \Gamma \,
\partial \Gamma$ term will cancel the second term on the right side of
that equation.

\subsubsection{{\em Non}-cancellation of the $\Gamma^3$ Terms}

While we are reasonably confident that, with these measures, $\Gamma \,
\partial \Gamma$ terms in the Bianchi identities will cancel, it is clear
that the $\Gamma^3$  terms will {\em not} cancel to machine accuracy.
These are all produced by connection of the double-$\Gamma$ terms in
the Riemann tensor.  However, when forming the Bianchi identities, the
connection takes place on an already-multiplied and summed $\Gamma \,
\Gamma$ product.  One cannot undo the sums and products, form averages,
and then re-form the triple-$\Gamma$ product.  And the product of averages
does {\em not} commute with the average of products.  The Bianchi
identities, therefore, will be left with terms proportional to the
truncation error and $(\partial g)^3$.  Unless some clever averaging
scheme can be found to allow the triple-$\Gamma$ terms to also cancel
to machine accuracy, any CT scheme developed along these lines will be
subject to truncation error.  The hope, then, is that cancellation of
second and first order derivatives of the connection (third and second
order derivatives of the metric coefficients) will be sufficient to
improve the stability of the discrete evolution scheme.

Therefore, while the staggered grid CT method clearly works in Riemann
normal coordinates, detailed numerical experiments will be needed to
see if it maintains its desirable properties when applied in a general
coordinate basis.

\subsection{Numerical Implementation}

\subsubsection{General Iterative Approach}

The goal of this CT scheme is to produce an interlocking, staggered
four-dimensional grid of ${\rm g}_{\alpha \beta}$ tensor values by successively
adding new half and whole temporal hypersurfaces.  Because interpolation
is in time, as well as space, the scheme necessarily will be an implicit
one and, therefore, iterative.  Our approach then will be to first
produce an initial guess for ${\rm g}_{\alpha \beta}$ on the next level of
hypersurfaces using an existing explicit scheme.  This solution will
not satisfy the interlocking staggered grid equations, so it will be
iterated using the latter until it does.  The exact iterative scheme is 
still under development, but one possible implementation is a multigrid 
technique in which the variables are the spatial ${\rm g}_{i j}$ and the 
equations are the ${{\rm R}^{i}}_{j} ~ = ~ 0$ field equations.  

This method appears similar to recently-suggested approaches in which
the constraints are re-solved at each time step.  There are, however,
two key differences.  First, the constraints are not solved explicitly.
Instead, the evolution equations are solved in such a way that they are
implicitly enforced.  This ensures a scheme in which the evolution and
constraints are fully compatible and not tracking different numerical
solutions to the differential equations.  Secondly, the equations
being iterated are implicit hyperbolic, not elliptic.  In addition to
information on the new hypersurface, at each iteration these equations
utilize information from the {\em previous} hypersurface --- information
that constraint equations do not have.  With this added information,
if implemented properly, the iterative scheme should exhibit faster
convergence than a regular constraint solver would have.

\begin{figure}
\begin{center}
\centerline{\psfig{figure=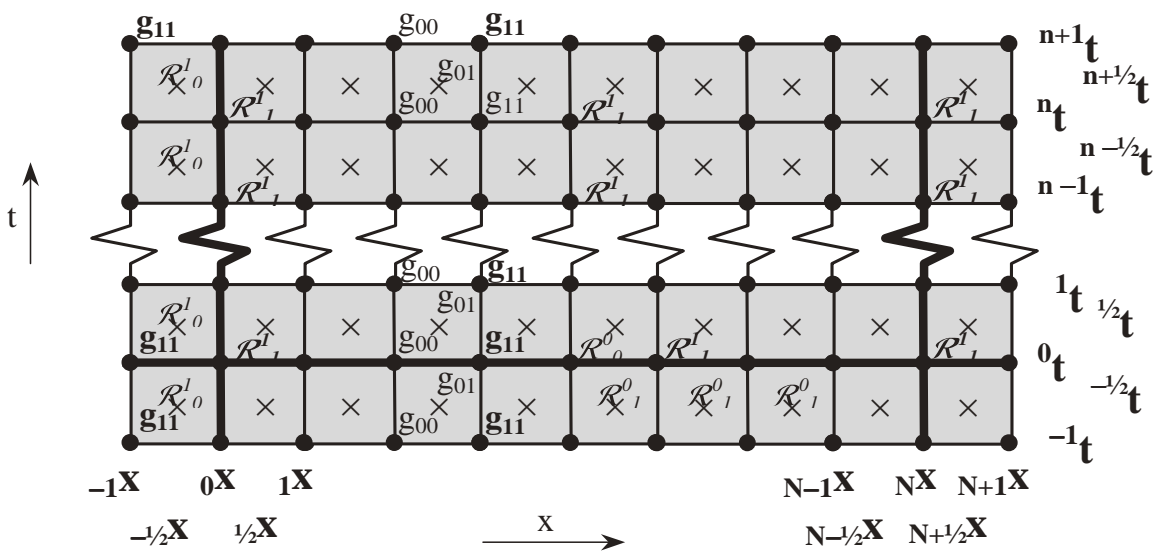,width=6.0in,angle=-90}}
\end{center}
\caption{Illustration of a 1+1 field evolution, including the 
initial data problem at $^0t$ and a one-point boundary data problem 
at $_0x$.  Bold quantities on the grid are solutions to the initial 
data and evolutionary problems.  The computation begins by using 
gauge conditions to 
specify ${\rm g}_{0 \, 0}$ at $^{-1}t$, $^0t$, and $^1t$ and ${\rm g}_{0 \, 1}$ 
at $^{-\frac{1}{2}}t$ and $^{\frac{1}{2}}t$. Then ${\rm {\rm R}^{0}}_1$, 
${\rm {\rm R}^{0}}_0$, and ${\rm {\rm R}^{1}}_1$ are solved for ${\rm g}_{1 \, 1}$ at 
$^{-1}t$, $^0t$, and $^1t$.  At the $_{0}x$ boundary ${\rm g}_{1 \, 1}$ 
may be specified, but at the $_{-1}x$ boundary ${\rm g}_{1 \, 1}$ 
must be consistent with ${\rm {\rm R}^{1}}_0 = 0$. 
Similarly, the ${\rm g}_{1 \, 1}$ at $_{N+1}x$ are 
obtained by solving ${\rm {\rm R}^{1}}_0 = 0$ at $_{N+\frac{1}{2}}x$. 
The evolution from $^{n}t$ to $^{n+1}t$ proceeds by specifying 
${\rm g}_{0 \, 1}$ at $^{n+\frac{1}{2}}t$ and ${\rm g}_{0 \, 0}$ at $^{n+1}t$.  
${\rm g}_{1 \, 1}$ is computed by solving ${\rm {\rm R}^{1}}_1 = 0$.  
The boundary conditions ${\rm {\rm R}^{1}}_0 = 0$ also must be applied at 
($^{n+\frac{1}{2}}t$, $_{-\frac{1}{2}}x$) and 
($^{n+\frac{1}{2}}t$, $_{N+\frac{1}{2}}x$) to obtain ${\rm g}_{1 \, 1}$ 
at $_{-1}x$ and $_{N+1}x$ on the new hypersurface.  
\label{fig5}}
\end{figure}

Figure \ref{fig5} shows implementation of a simple 1+1 space-time
problem.  Comparison of this figure with Figure \ref{fig4} will give
insight into implementation of the full 4-D interlocking grid.  $^n{\rm g}_{0
0}$ and $^n{\rm g}_{1 1}$ are quantities known from the previous time step,
the first being a gauge condition and the second being the solution.
$^{n+\frac{1}{2}}{\rm g}_{1 0}$ and $^{n+1}{\rm g}_{0 0}$ are gauge conditions on
the {\em new} hypersurfaces, and $^{n+1}{\rm g}_{1 1}$ is the new solution
that will be determined by the iterative numerical scheme.

\subsubsection{Using Gauge Conditions and Time Stepping}

Specification of the coordinate gauge is similar to doing so in classic
3+1 schemes: ${\rm g}_{0 0}$ is freely specified at cube corners on whole
hypersurfaces, and ${\rm g}_{0 i}$ is freely specified at cube edge centers on
half hypersurfaces.  ${\rm g}_{0 i}$ is related to the shift vector $\beta^{i} =
{\rm g}^{0 i}$, and both are properly located at half time steps vector points.
${\rm g}_{0 0} = -\alpha^2 + {\rm g}_{i j} \beta^{i} \beta^{j}$ is related to the
lapse $\alpha$, and properly located in this scheme at scalar points.
The ${\rm g}_{i j}$ are the six unique gravitational potential fields that we
intend to solve.

A typical time step then begins with ${\rm g}_{\alpha \beta}$ components known
on hypersurfaces $^{n-1}t$, $^{n-\frac{1}{2}}t$, and $^nt$.  One then
specifies the ${\rm g}_{0 i}$ vector on hypersurface $^{n+\frac{1}{2}}t$ and
the ${\rm g}_{0 0}$ scalar at $^{n+1}t$.  The explicit predictors for ${\rm g}_{i
j}$ at $^{n+1}t$ then complete the new ${\rm g}_{\alpha \beta}$ field, and
the iteration on the ${\rm g}_{i j}$ can then begin.  If the ${\rm g}_{0 \alpha}$
are dependent on the ${\rm g}_{i j}$ field (which generally will be the case
here), then the gauge conditions need to be updated at each iteration
for a consistent solution.

\subsubsection{Constructing the Inverse Metric}

The inverse of ${\rm g}_{\alpha \beta}$ is often needed to raise the connection
coefficients for use in the evolution equations.  In principle these
${\rm g}^{\alpha \beta}$ should also be located at second-ranked tensor
grid points.  In practice, however, we never need the actual staggered
${\rm g}^{\alpha \beta}$ fields.  Instead, we need their interpolated averages
$\bar{\rm g}^{\alpha \beta}$ at third-ranked tensor points.  Furthermore,
in order for the raising and lowering of connection coefficient indices
to commute, the staggered field of ${\rm g}^{\alpha \beta}$ values must be
constructed in such a way that its average, and that of ${\rm g}_{\alpha
\beta}$, are orthogonal at those third-ranked tensor points
\begin{eqnarray}
\bar{\rm g}^{\alpha \mu} \, \bar{\rm g}_{\mu \beta} & = & {\delta^{\alpha}}_{\beta}
\end{eqnarray}
So construction of the index-raising tensor $\bar{\rm g}^{\alpha \beta}$
is a straightforward matter of averaging ${\rm g}_{\alpha \beta}$ to places
where the $\Gamma_{\alpha \beta \gamma}$ are computed, and then inverting
that average {\em locally} at those grid points.  The actual staggered
fields of ${\rm g}^{\alpha \beta}$ values, whose averages should give us these
$\bar{\rm g}^{\alpha \beta}$ at third-ranked tensor points, never need to
be determined.  This procedure is fast, gives us raising and lowering
operators that commute, and follows the afore-mentioned rule of averaging
first and then multiplying and summing second.

\subsubsection{The Initial Data Problem}
 
The initial data problem for a staggered grid will be a little more
complicated than that for a non-staggered scheme.  In the latter case,
there are 12 unknowns (${\rm g}_{i \, j}$ and ${\rm g}_{i \, j \, , \, 0}$) and four
equations (${{\rm R}^{ 0 }}_{ \beta } = 0$) on the initial hypersurface at $t
= \, ^0t$, for a net total of eight degrees of freedom.  In the staggered
case, ${{\rm R}^{0}}_{0}$ is computed at $t = \, ^0t$, but the ${{\rm R}^{ 0 }}_{
j }$ are computed at $t = \, ^{-\frac{1}{2}}t$.

Solution of the momentum constraints, given ${\rm g}_{i \, j}$ at $t = \, ^0t$,
is fairly straightforward in the staggered case.  They are not functions
of second-order time derivatives of the metric ({\it e.g.}, ${\rm g}_{i \,
j \, , \, 0 \, 0}$), so the placement of those constraints at $t = \,
^{-\frac{1}{2}}t$ allows them to directly relate $^{-1}{\rm g}_{i \, j}$ at to
$^0{\rm g}_{i \, j}$.  All $\Gamma$s can be computed in a staggered manner.
The momentum constraint solution at $^{-\frac{1}{2}}t$, then, would
have six unknowns ($^0{\rm g}_{i \, j}$) and three constraints, leaving three
degrees of freedom, just like the non-staggered grid case.

The Hamiltonian constraint, however, presents a problem.  While
${{\rm R}^{0}}_{0}$ also does not involve any second-order time derivatives,
nevertheless it does involve {\em first} order time derivatives ${\rm g}_{i
\, j \, , \, 0}$.  Those still are defined on half-hypersurfaces and,
therefore, need to be {\em interpolated} to $^0t$.  That is, we need the
${\rm g}_{i \, j \, , \, 0}$ at both $^{-\frac{1}{2}}t$ and $^{\frac{1}{2}}t$
anyway, even if we are not going to compute ${\rm g}_{i \, j \, , \, 0 \, 0}$.
At first glance there does not appear to be a method of providing an
accurate ${\rm g}_{i \, j \, , \, 0} (t)$ field to properly compute this
interpolation.  Of course, one simply could extrapolate ${\rm g}_{i \, j \,
, \, 0}$ at $^{-\frac{1}{2}}t$ forward to $^0t$ ({\it i.e.}, assume ${\rm g}_{i
\, j \, , \, 0 \, 0} = 0$) and then solve ${{\rm R}^{0}}_{0} = 0$ there.
The Hamiltonian constraint solution then would have the six ${\rm g}_{i \,
j}$ unknowns and one constraint --- five degrees of freedom --- just
like the non-staggered case.

But a serious problem still remains.  {\em There is no means of enforcing
the Hamiltonian constraint at $t = \, ^1t$.}  While the staggered grid
is, in principle, capable of doing that, it can do so only through
the evolution equations ${{\rm R}^{i}}_{j} = 0$.  But at this stage we
have not yet begun to enforce the evolution.  One possible method of
solving this is to do nothing.  Just accept the fact that, at $t =
\, ^1t$, ${{\rm R}^{0}}_{0} = 0$ is good only to truncation accuracy.
Choosing a very small $\Delta t$ (a ``thin sandwich'') would keep
this error small.  A second approach would be to also explicitly
enforce the constraint on $t = \, ^1t$, which would reduce the number
of degrees of freedom of the Hamiltonian problem from five to four.
While producing a more constrained problem, this solution would result
in two successive hypersurfaces on which the Hamiltonian constraint is
satisfied.  Yet a third alternative would be to locate ${{\rm R}^{ 0 }}_{
j }$ on the $^{\frac{1}{2}}t$ hypersurface and extrapolate quantities
like ${\rm g}_{i \, j \, , \, 0}$ {\em forward} to $^1t$.  The problem with
this approach is that, while the Hamiltonian constraint is satisfied for
the {\em extrapolated} ${\rm g}_{i \, j \, , \, 0}$ field, when the evolution
is begun, the Hamiltonian constraint that is implicit in the evolution
equations will use fields that are {\em interpolated} in time between
$^{-\frac{1}{2}}t$ and $^{\frac{1}{2}}t$.  There will be, therefore, an
implicit constraint violation injected into the evolution at the outset.

The elegant, and proper, method of solving this problem is to {\em
solve all ten of the Einstein field equations simultaneously on the
initial hypersurfaces}.  There will be 18 unknowns (${\rm g}_{i \, j}$ at
$^{-1}t$, $^0t$, and $^1t$) and 10 equations, leaving 8 degrees of
freedom, just like the non-staggered case.  The resulting fields will
be properly staggered, and the $\Gamma$s will be properly staggered and
interpolated.  The Hamiltonian constraint will be satisfied at $^0t$
using the correctly interpolated fields, and it will be satisfied at
$^1t$ (and even $^{-1}t$) as well, because the evolution equations
(and therefore the Bianchi identities) are fully enforced.  Similarly,
the momentum constraints will be satisfied at $^{-\frac{1}{2}}t$
and $^{\frac{1}{2}}t$ for the same reason, regardless of whether they
are actually applied at $^{-\frac{1}{2}}t$ or at $^{\frac{1}{2}}t$.
The reader will, of course, recognize that this is more than solving an
initial data problem; in actuality the proposed scheme solves the initial
data problem plus the first file evolutionary time step simultaneously.
This is done to ensure that the initial data on the first three 
hypersurfaces are solutions of the discrete, staggered evolutionary 
field equations.  No constraint violation will be introduced implicitly 
other than what is naturally present in the evolutionary method already.

\subsubsection{Boundary Conditions}

\subsubsubsection{One-point Boundary Conditions}

When $\bf{g}$ and $\bf{n} \cdot \nabla \bf{g}$ are specified on the
same boundary, where $\bf{n}$ is the boundary normal, the boundary data
problem is similar to the initial data problem.  In all such cases,
the boundary constraints are given by
\begin{eqnarray}
{\rm n}_{\mu} \, {{\rm R}^{ \mu }}_{ \beta } & = & 0
\end{eqnarray}
and for rectilinear grids with the boundary normal being a coordinate
unit 1-form
$\bf{n} = \bf{w}^{(i)}$, this yields
\begin{eqnarray}
\label{boundary_constraints}
{{\rm R}^{ ( \, i \, ) }}_{ \beta } & = & 0
\end{eqnarray}
where the symbol $(i)$ is a label indicating the boundary direction in
question, not strictly a coordinate index.  The equation ${{\rm R}^{ ( \,
i \, ) }}_{ i }$ (no sum) is located at 3-cube corners and plays the
role of boundary constraint in much the same manner as the Hamiltonian
constraint does at $^0t$.  Similarly, the equations ${{\rm R}^{ ( \, i \, )
}}_{ a } = 0$ ($a \ne i$) play the same role as the momentum constraints
did earlier.  By analogy, then, the boundary problem is as follows.
There are 12 unknowns (${\rm g}_{a \, b}$ [ $a \ne i$, $b \ne i$] at $x^{(i)}
= \, _0x^{(i)}$ and at $x^{(i)} = \, _{-1}x^{(i)}$, {\it i.e.}, on the
$i^{\rm th}$ boundary and one ghost node {\em beyond} the boundary),
and there are 4 equations (\ref{boundary_constraints}).  This leaves
8 degrees of freedom again, which must be specified with additional
boundary conditions.  The diagonal constraint ${{\rm R}^{ ( i ) }}_{ i } =
0$ is applied at $_0x^{(i)}$ at 3-cube corners.

Two of the off-diagonal constraints (${{\rm R}^{ ( \, i \, ) } }_{
j } = 0$ $[j \ne i]$) are applied on whole hypersurfaces at
$_{-\frac{1}{2}}x^{(i)}$, and the final constraint ${{\rm R}^{ ( \,
i \, ) }}_{ 0 } = 0$ is applied also at $_{-\frac{1}{2}}x^{(i)}$
but on time half-hypersurfaces.  The reader will note that
this latter set of equations is related to the set of momentum
constraints ${{\rm R}^{ ( 0 ) }}_{ i } = 0$, but the momentum constraints
are defined at
$_{\frac{1}{2}}x^{(i)}$, $_{\frac{3}{2}}x^{(i)}$, $_{\frac{5}{2}}x^{(i)}$,
..., $_{N+\frac{1}{2}}x^{(i)}$ and propagated forward by the evolution
equations.  The constraint at $_{-\frac{1}{2}}x^{(i)}$ must be explicitly
enforced in most cases, along with the three others and the eight
freely-specified ${\rm g}_{\alpha \beta}$ at $_{-1}x^{(i)}$  and $_{0}x^{(i)}$.

\subsubsubsection{Two-point Boundary Conditions}

Any number of additional boundary data problems are possible.  ${{\rm R}^{
( \, i \, ) }}_{ a } = 0$ could be specified at the upper $(i)$ boundary,
while ${{\rm R}^{ ( \, i \, ) }}_{ i } = 0$ could be be specified at the lower
$(i)$ boundary, for example.  Also, some of the eight free ${\rm g}_{\alpha
\beta}$ could be specified on opposing boundaries as well.

For periodic boundary conditions, the six unknown ${\rm g}_{\alpha \beta}$
at $_{-1}x^{(i)}$  are set to those near $_{N}x^{(i)}$, and those at
$_{N+1}x^{(i)}$ are set to those near $_{0}x^{(i)}$.  No constraints are
applied explicitly, only implicitly through the wrapping conditions and
the evolutionary solution of the field equations.

\subsubsection{Implementation Summary}

It is useful to summarize how a complete problem will proceed.  The
computation begins by using gauge conditions to specify ${\rm g}_{0 \, 0}$
at $^{-1}t$, $^0t$, and $^1t$ and ${\rm g}_{0 \, i}$ at $^{-\frac{1}{2}}t$
and $^{\frac{1}{2}}t$.  The full initial data plus time-step problem, 
including the
field equations, is then solved for ${\rm g}_{i \, j}$ on $^{-1}t$, $^0t$, and
$^1t$, applying eight freely-specified ${\rm g}_{i \, j}$ (or eight functions
thereof) in the process.  Appropriate boundary constraints also need to
be applied in order to obtain a consistent solution.

The field equations are generated as follows.  An initial guess for
the ${\rm g}_{i \, j}$ is obtained by some means, perhaps using conformal or
other existing initial value methods or, for the evolution, an explicit
forward integration scheme.  The full ${\rm g}_{\alpha \beta}$ field values
then are differenced onto third-rank tensor grid points, and the
$\Gamma_{\alpha \beta \gamma}$ are formed.  The ${\rm g}_{\alpha \beta}$
values are also averaged to those same third-rank tensor points, and
an inverse of that average $\bar{\rm g}^{\alpha \beta}$ is used to raise
the connection coefficients (equations \ref{singly_raised_Gamma}
and \ref{doubly_raised_Gamma}).  The $\Gamma$s, in turn, then are
differenced and averaged to second-rank tensor points for the Riemann
tensor calculation.  (Note that there will be no need for values at other
fourth-rank tensor points [hypercube body centers], as Riemann will be
immediately contracted into Ricci.)  Equation (\ref{mixed_riemann})
should suffice for the Riemann calculation ({\it i.e.}, the explicit
version [equation \ref{new_mixed_riemann}] should not be needed), because
our raising and lowering operators commute, rendering the results of
equations (\ref{singly_raised_Gamma}) and (\ref{alt_raised_Gamma}) the
same to machine accuracy.  Contraction to Ricci is trivial, as it sums
Riemann components already computed at second-rank tensor points, and
the ${{\rm R}^{i}}_{j}$ are then tested to see if they are zero.  If not, the
local values of ${\rm g}_{i \, j}$ are modified in an appropriate manner, and
the computation of the Ricci components is repeated until convergence.
It is important to remember that at grid boundaries, the boundary
constraints (which are evolutionary equations in their own right) must
be applied and iterated upon as well, and the freely-specified boundary
conditions must be applied.

When the iterative solution is acceptable, the computation continues to
the next hypersurface, first specifying the gauge conditions and then
solving the field equations.

\section{Discussion}

The largest uncertainty in the proposed method is the effect of the
truncation errors introduced by non-cancellation of the triple-$\Gamma$
terms.  Analytically we can be sure of exact constraint propagation
only if the Bianchi identities are satisfied exactly, and that is not
the case in a general coordinate basis.  On the other hand, the fact
that the method does work in a local Riemann normal system, and has some
attractive properties of a finite differential calculus, are encouraging.
Furthermore, most of the time-dependent derivatives of ${\bf g}$
do cancel in the general case, and this should enhance stability.
Analytical investigation of the stability of this method is difficult,
so we have chosen to do so numerically.

Two numerical implementations of this scheme are being developed
currently.  The first assumes symmetries in two spatial dimensions,
rendering the scheme explicit and avoiding the need for iteration.
The second is a full implementation using staggered grids in four
dimensions.  Results of these studies will help in determining the
stability of the method.

At best, the method is expected to be conditionally stable.  The
Evans-Hawley and Yee methods have this property, and are subject to a
Courant-like condition on the time step \cite{deraedt02}.  With 
$c \, = \, 1$, this implies $\Delta t \, < \, \Delta x$, which a 
typical condition applied in most numerical relativity implementations.

If it can be shown that such techniques provide stable constraint
transport for finite difference implementations of numerical relativity,
then other implementations might benefit from similar schemes that use 
four-dimensional grids and that have temporal and spatial derivatives 
that commute.  In the finite element case, this would begin by extending 
the elements into the time direction, rather than simply time-stepping a 
three-dimensional finite element grid.  However, some additional features
would have to be introduced, analogous to grid staggering, to ensure
commutation of spatial and time derivatives across element boundaries.

The pseudo-spectral may be intractable at the present time.  As spatial 
derivatives are computed using the full spatial extent of the grid, in 
order to create a method in which these commuted with time derivatives, 
one could imagine using many, if not all, previous and future time 
hypersurfaces to compute the latter.  This would be a truly four-dimensional 
grid method and involve solving the entire spacetime structure in one giant 
iterative procedure.  Present-day computers still struggle with three-dimensional
explicit schemes, so a four-dimensional implicit one is clearly beyond
current technology.  However, in the not-too-distant future such codes
might begin to be feasible.


\begin{acknowledgments}
The author has benefited greatly from discussions with F. Estabrook, L.
Lindblom, M. Miller, D. Murphy, and M. Scheel.  He is also grateful for a
JPL Institutional Research and Development grant, and for the continued
hospitality of the TAPIR group at Caltech.  This research was performed
at the Jet Propulsion Laboratory, California Institute of Technology,
under contract to the National Aeronautics and Space Administration.
\end{acknowledgments}

\clearpage



\end{document}